\documentclass[12pt]{article}
\usepackage{graphicx}
\usepackage{booktabs}
\usepackage{multirow}
\usepackage[numbers]{natbib}
\usepackage{centernot}
\usepackage{xcolor}
\usepackage{amssymb}% http://ctan.org/pkg/amssymb
\usepackage{pifont}% http://ctan.org/pkg/pifont
\usepackage[normalem]{ulem}

\renewcommand{\baselinestretch}{1.5}
\newcommand {\bd}[1]{\mbox{\boldmath$#1$}}

\newcommand {\rb}[1]{\raisebox{1.5ex}[0pt]{#1}}

\def\logit{{\mathrm{logit}}}
\def\bSig\mathbf{\Sigma}

\usepackage{chngcntr}
\counterwithout{figure}{section}

\begin{document}

\title{\bf Estimating the treatment effect for adherers using multiple imputation}

\author{ 
{\bf Junxiang Luo } \\ \small 200 Technology Square, Cambridge, MA 02139, USA \\
\small Email: junxiang.luo@Modernatx.com \\ {\bf Stephen J. Ruberg } \\ \small Analytix Thinking, LCC, 11121 Bentgrass Court, Indianapolis, IN 46236, USA \\ \small Email: AnalytixThinking@gmail.com \\
{\bf Yongming Qu* } \\ \small Department of Statistics, Data and Analytics, Eli Lilly and Company, Indianapolis, IN 46285, USA \\ \small Email: qu\_yongming@lilly.com \\
}
\maketitle
\noindent
*Correspondence: Yongming Qu, Department of Statistics, Data and Analytics, Eli Lilly and Company, Lilly Corporate Center, Indianapolis, IN 46285, U.S.A. Email: qu\_yongming@lilly.com.

\label{firstpage}
\newpage
%  put the summary for your paper here

\begin{abstract}
Randomized controlled trials are considered the gold standard to evaluate the treatment effect (estimand) for efficacy and safety. According to the recent International Council on Harmonisation (ICH)-E9 addendum (R1), intercurrent events (ICEs) need to be considered when defining an estimand, and principal stratum is one of the five strategies to handle ICEs. Qu et al. (2020, Statistics in Biopharmaceutical Research 12:1-18) proposed estimators for the adherer average causal effect (AdACE) for estimating the treatment difference for those who adhere to one or both treatments based on the causal-inference framework, and demonstrated the consistency of those estimators; however, this method requires complex custom programming related to high-dimensional numeric integrations. In this article, we implemented the AdACE estimators using multiple imputation (MI) and constructs CI through bootstrapping. A simulation study showed that the MI-based estimators provided consistent estimators with the nominal coverage probabilities of CIs for the treatment difference for the adherent populations of interest. As an illustrative example, the new method was applied to data from a real clinical trial comparing 2 types of basal insulin for patients with type 1 diabetes.
\end{abstract}

%  Please place your key words in alphabetical order, separated
%  by semicolons, with the first letter of the first word capitalized,
%  and a period at the end of the list.
%

\maketitle
\newpage

%\begin{abstract}
%\noindent
%{\small SUMMARY.  }
%\end{abstract}

%\noindent 
%{\small
%{\bf KEY WORDS} Information gain, Kullback-Leibler, Likelihood reduction factor, Treatment effect, Surrogate marker.
%}
\section{Introduction}
Choosing and defining estimands and constructing corresponding estimators are integral parts of randomized controlled clinical trials. The International Council on Harmonisation (ICH) provides a general framework for choosing and defining estimands using a few key attributes: treatment of interest, population, handling intercurrent events (ICEs), endpoint, and population-level summary \cite{ich2019e9}. There are three possible populations in defining estimands: the whole targeted population, a subset of population based on baseline covariates, and a principal stratum defined by post-baseline variables \citep{scharfstein2019constructive, qu_lipkovich2021}. A principal stratum is a subset of the study population defined by the potential outcome of one or more post-randomization variables \citep{air1996identification, imbens1997bayesian, frangakis2002principal}. 
The most used estimands in clinical trials are to estimate the treatment effect for the whole targeted study population. 
The whole targeted study population in defining estimands is widely used as it theoretically maintains randomization, but the difficulty arises when, as inevitably happens, some patients do not provide complete efficacy response data relevant to the study objective due to intercurrent events. ICH E9 (R1) defines ICEs as “Events occurring after treatment initiation that affect either the interpretation or the existence of the measurements associated with the clinical question of interest. It is necessary to address intercurrent events when describing the clinical question of interest in order to precisely define the treatment effect that is to be estimated.” Deciding which strategy to use in handling ICEs can lead to different estimands. This is central to the development of ICH E9 (R1) which provides a framework and a language for defining the estimand first, and subsequently, an appropriate data handling and analysis approach.  
As mentioned in ICH E9 (R1), the definition of the estimand should be guided by the study objective. Deciding the treatment effect for the whole study population is an important question, but it may not be the only or the most important question. The authors of the recent research \cite{ruberg2017considerations, qu2020general, qu2021implementation} provide arguments of importance of the treatment effect for principal strata of adherers, defined as the population of patients without ICEs. The treatment effect for the whole study population answers the question, ``what is the overall expected treatment effect that would be communicated to an individual patient before that patient takes the medication?", which is an \emph{unconditional expectation} regardless of adherence status. 
%Mathematically, it can be expressed as
%\[ E[Y(1)-Y(0)], \]
%where $Y(1)$ and $Y(0)$ denote the p
The treatment effect for the adherers answers the question, ``what is the treatment effect in a patient who can adhere to the treatment?", which is a {\em conditional expectation} regarding adherence status.
%\[
%E[Y(1)-Y(0)|S].
%\]

The estimation of the treatment effect for a principal stratum has attracted interest for data analysis in clinical trials previously \citep{mehrotra2006comparison, shepherd2008does, permutt2018effects, bornkamp2020estimating, bornkamp2021principal}. 
There are a variety of approaches for constructing estimators of the treatment effect in a principal stratum. Most methods require the {\em monotonicity} assumption and/or using the principal score \citep{bornkamp2020estimating, zhang2003estimation, hayden2005estimator, small2007stochastic, jin2008principal, jo2009use, chiba2011simple, ding2011identifiability, vanderweele2011principal, lu2013rank, feller2016principal, magnusson2018bayesian}.
%The first type of methods rely on the {\em monotonicity} assumption \citep{bornkamp2020estimating, zhang2003estimation, ding2011identifiability, lu2013rank, magnusson2018bayesian, chiba2011simple, vanderweele2011principal, jin2008principal}. 
Monotonicity basically assumes the potential outcome  for the stratification indicator is a monotone function of the treatment indicator. The monotonicity assumption imposes a deterministic relationship on the potential outcomes (random variables) of the principal stratum variable(s). Qu et al. \cite{qu2020monotonicity} demonstrate the implausibility of such an assumption in many situations from a theoretic perspective and illustrate this implausibility using a real data example in a cross-over study. The principal score is the probability of a subject belonging to the principal stratum, modeled via baseline covariates. As an apparent drawback, methods based on the principal score assume the principal stratum can be fully modeled through baseline covariates.   
%The second type of methods focus on estimating the bounds of a treatment effect on a principal stratum or estimating such a treatment effect via introducing some sensitivity parameters (not estimable from data alone) \citep{mehrotra2006comparison, imai2008sharp,shepherd2011sensitivity,lou2019estimation}. These methods generally can only serve as sensitivity analyses since they cannot provide consistent point estimators.  
%The third type of methods estimate the treatment effect for a principal stratum using the principal score, the probability of a subject belonging to the principal stratum, via baseline covariates, with or without the monotonicity assumption \citep{hayden2005estimator,jo2009use, small2007stochastic, feller2016principal,bornkamp2020estimating}. Some other methods estimate the treatment effect using post-baseline intermediate outcomes. 
Recently, Louizos et al. \cite{louizos2017causal} proposed methods directly estimating the potential outcome of the response variable under the alternative treatment if the principal stratum can be observed in one treatment group. 
Qu et al. \cite{qu2020general} developed estimators for the adherer average causal effect (AdACE) based on the causal inference framework for the treatment effect for those who can adhere to one or both treatments by modeling the potential outcome of the response variable and/or the principal score via baseline covariates and potential post-baseline intermediate measurements. It is important to include post-baseline intermediate measurements since patients and their treating physicians in clinical trials most often make decisions about whether or not to adhere to the randomized study treatment based on their efficacy and safety responses to the assigned treatment. Details on the implementation of estimators for the AdACE are given in Qu et al. \cite{qu2021implementation}. Barriers to a wide application of these estimators include the complex estimation process, the time-consuming computation, and the requirement of customized programs for individual clinical trials. 

Multiple imputation (MI), proposed by Rubin \cite{rubin1987multiple}, is widely used in handling missing values and could be an alternative approach to construct estimators for the AdACE. The advantage of using MI is that the estimators can easily be calculated based on the imputed potential outcomes. 
In this article, we propose using MI to construct the estimators for the AdACE. The advantage of this approach is that it can utilize the existing estimation procedures for ``complete" data after MI to estimate treatment effect in the stratum of interest. 
In Section 2, we will review the theoretical framework and outline the process of the MI-based estimators for adherers. The definition of adherence may be study specific, but generally we consider a patient to be adherent if the patient predominantly takes their estimand-defined study treatment (e.g., no intercurrent events) throughout the intended duration of the trial. As we are interested to estimate the potential outcome under the intended treatment regimen for adherers, the (potential) outcomes under a treatment regimen deviating away from the randomized treatment (e.g., treatment discontinuations or treatment switch) are not relevant in our estimation. In Section 3, simulations are conducted to evaluate the performance of the MI-based estimator. In Section 4, the application of the MI-based estimator is illustrated with a real clinical study. Finally, Section 5 serves as the summary and discussion. 

\section{Methods}
\label{sec:methods}
Let $(\bd X_{ij}, \bd Z_{ij}, Y_{ij}, \bd I_{ij})$ denote the data for assigned treatment $i\;(i=0,1)$ and subject $j \; (1 \le j \le n_i)$, where $\bd X_{ij}$ is a vector of baseline covariates, $\bd Z_{ij} = (Z_{ij}^{(1)}, Z_{ij}^{(2)}, \ldots, Z_{ij}^{(K-1)})'$ is a vector of intermediate repeated measurements, $\bd I_{ij} = (I_{ij}^{(1)}, I_{ij}^{(2)}, \ldots, I_{ij}^{(K-1)})'$, and  $I_{ij}^{(k)}$ ($1 \le k \le K-1$) is the indicator variable for whether a patient is adherent to treatment after intermediate time point $k$. Note $Z$ can be intermediate measurements of the same variable as $Y$, or intermediate outcomes of other ancillary variables, or include both.
Then, the adherence indicator for the study treatment is given by:
\[
A_{ij} = \prod_{k=1}^{K-1} I_{ij}^{(k)}.
\]  
We use ``$(t)$" following the variable name to denote the potential outcome under the hypothetical treatment $t \; (t=0,1)$. For example, $Y_{ij}(t)$ denotes the potential outcome for subject $j$ randomized to treatment $i$ if taking treatment $t$. Generally, $Y_{ij}(i)$ can be observed but $Y_{ij}(1-i)$ cannot be observed in parallel studies. 

Four principal strata were discussed in Qu et al. \cite{qu2020general}: 
\begin{list}{$\bullet$}{\leftmargin=0cm \itemindent=1.5cm}
    \item The whole study population: $S_{**}=\{(i,j): A_{ij}(0) \in \{0,1\}, A_{ij}(1) \in \{0,1\}\}$. 
    \item Patients who can adhere to the experimental treatment: $S_{*+}=\{(i,j): A_{ij}(1)=1 \}$
    \item Patients who can adhere to the control treatment: $S_{+*}=\{(i,j): A_{ij}(0)=1 \}$
    \item Patients who can adhere to both treatments: $S_{++}=\{(i,j): A_{ij}(0)=1, A_{ij}(1)=1\}$
\end{list}

Qu et al. \cite{qu2020general} provides estimators for $S_{*+}$, $S_{+*}$, and $S_{++}$ under the following assumptions:
\begin{list}{$\square$}{\leftmargin=0em \itemindent=2cm}
	\item[A1:] $Y=Y(1)T+Y(0)(1-T)$
	\item[A2:] $Z = Z(1)T+Z(0)(1-T)$
	\item[A3:] $A=A(1)T + A(0)(1-T)$
	\item[A4:] $T \perp \{Y(1), A(1), Z(1) ,Y(0), A(0), Z(0) \} | X$
	\item[A5:] $A(i) \perp \{Y(1), Y(0), Z(1-i)\} | \{X, Z(i)\}, \quad \forall i=0,1 $
	\item[A6:] $Y(i) \perp Z(1-i) | \{X, Z(i)\}, \quad \forall i=0,1$
	\item[A7:] $Z(0) \perp Z(1)|X$
\end{list}
The estimators provided in Qu et al. \cite{qu2020general} were rather complex; however, the idea is to estimate the potential response $Y_{ij}(1-i)$ and/or the potential adherence status $A_{ij}(1-i)$ under the alternative treatment. 

Alternatively, estimators can be achieved naturally with the approach of MI. For a subject $j$ in assigned treatment $i$, the potential outcome $Y_{ij}(1-i)$, $\bd Z_{ij}(1-i)$, and $\bd I_{ij}(1-i)$ are unobserved for the alternative treatment $1-i$ and can be imputed using a model estimated from all data from treatment $1-i$, i.e., $\{(\bd X_{1-i,j}, \bd Z_{1-i,j}, Y_{1-i,j}, \bd I_{1-i, j}): 1 \le j \le n_{1-i} \}$, and his/her own baseline value $\bd X_{ij}$. Let $\{ (\bd Z_{ij}(1-i)^{(m)}, Y_{ij}(1-i)^{(m)}, \bd I_{ij}(1-i)^{(m)}), 1 \le m \le M \}$ be the $M$ imputed values for the potential outcomes under treatment $1-i$ for subjects assigned to treatment $i$. Note the unobserved outcomes for the potential treatment $i$ for those assigned to treatment $i$ (due to random missingness or non-adherence) can also be simultaneously imputed, denoted by $\{ (\bd Z_{ij}(i)^{(m)}, Y_{ij}(i)^{(m)}, 1 \le m \le M \}$.  
The potential adherence indicator under the alternative treatment based on the imputed values is calculated as:
\[
A_{ij}(1-i)^{(m)} = \prod_{k=1}^{K-1} I_{ij}(1-i)^{(k,m)}.
\]
\begin{table}[h!tb] \centering
\caption{Illustration of MI to impute the potential outcome under treatment $T=t$ for patients assigned to treatment $T=1-t$}
\begin{tabular}{cccccccccc}
\hline\hline
Subject & Randomized Treatment & $X$ & $Z^{(1)}$ & $Z^{(2)}$ & $Z^{(3)}$ & $I^{(1)}$ & $I^{(2)}$ & $I^{(3)}$ & $Y$ \\
001 & $t$ & $\checkmark$ & $\checkmark$ & $\checkmark$ & $\checkmark$& 1 & 1 & 1 & $\checkmark$\\
002 & $t$ & $\checkmark$ & $\checkmark$ & $\checkmark$ & $\checkmark$ & 1 & 1 & 0 & $\cdot$ \\
003 & $t$ & $\checkmark$ & $\checkmark$ & $\checkmark$ & $\cdot$& 1 & 0 & 0 & $\cdot$ \\
$\cdots$ &&&&&&&&&\\
101 & $1-t$ & $\checkmark$ & $\cdot$ & $\cdot$ & $\cdot$ & $\cdot$ & $\cdot$ & $\cdot$ & $\cdot$\\
102 & $1-t$ & $\checkmark$ & $\cdot$ & $\cdot$ & $\cdot$& $\cdot$ & $\cdot$ & $\cdot$ & $\cdot$ \\
103 & $1-t$ & $\checkmark$ & $\cdot$ & $\cdot$ & $\cdot$& $\cdot$ & $\cdot$ & $\cdot$ & $\cdot$ \\
\hline\hline
\multicolumn{10}{p{15cm}}{Abbreviations: MI, multiple imputation; ``$\checkmark$", observed data; ``$\cdot$", unobserved data. 
} 
\end{tabular}
\label{table_MI}
\end{table}

After imputations, for each patient the (potential) outcome $Y$ and adherence $A$ under both treatments are available. Then, the mean response for each treatment can be calculated by simply taking the average of the (potential) outcome of $Y$ for the (potential) adherers.
The estimation of the treatment effect for the whole study population using MI has been extensively studied in the literature \cite{mallinckrodt2016analyzing}, so we will not discuss it here. 
Essentially, the estimators for $S_{*+}$ and $S_{+*}$ can be constructed in the exact same way by symmetry. Therefore, we only provide the estimators for the mean response in each treatment on populations $S_{*+}$ and $S_{++}$  (Table \ref{tab:estimators}), which are most relevant for placebo-controlled trials and active-comparator trials, respectively, as argued by Qu et al. \cite{qu2020general}. 
For each treatment, the estimator can be constructed using patients randomly assigned to treatment $T=0$, $T=1$, and all patients. Generally, it is preferable to use all patients in constructing the estimators as this most closely represents the study population. The treatment difference can be calculated easily based on the estimators for individual treatments. 

\begin{table}[h!tb] \centering
\caption{Estimators for the mean response on principal strata defined by treatment adherence}
\begin{tabular}{cccc}
        \hline\hline
        PS & Treatment & Patient & Estimator \\ \hline
&  & $E_0$ & $\frac{1}{M} \sum \limits_{m=1}^M \left\{ \frac{\sum_{j=1}^{n_0}  A_{0j}(1)^{(m)} \left( A_{0j}Y_{0j}+(1-A_{0j})Y_{0j}(0)^{(m)} \right) } {\sum_{j=1}^{n_0} A_{0j}(1)^{(m)} } \right\}$ \\
 & $T=0$  & $E_1$ &
$\frac{1}{M} \sum\limits_{m=1}^M \left\{ \frac{\sum_{j=1}^{n_1}  A_{1j} Y_{1j}(0)^{(m)} } {\sum_{j=1}^{n_1} A_{1j}} \right\}$ \\
{$S_{*+}$}  &   & $E_0\cup E_1$ & $\frac{1}{M} \sum \limits_{m=1}^M \left\{ \frac{\sum\limits_{j=1}^{n_0}  A_{0j}(1)^{(m)} \left( A_{0j}Y_{0j}+(1-A_{0j})Y_{0j}(0)^{(m)} \right) + 
 \sum\limits_{j=1}^{n_1}  A_{1j} Y_{1j}(0)^{(m)}
 } {\sum\limits_{j=1}^{n_0} A_{0j}(1)^{(m)} + \sum\limits_{j=1}^{n_1} A_{1j}} \right\}$ \\ \cline{2-4}
&  & $E_0$ & $\frac{1}{M} \sum \limits_{m=1}^M \left\{ \frac{\sum_{j=1}^{n_0}  A_{0j}(1)^{(m)} Y_{0j}(1)^{(m)} } {\sum_{j=1}^{n_0} A_{0j}(1)^{(m)}} \right\}$ \\
 & $T=1$ & $E_1$ &  $ \frac{\sum_{j=1}^{n_1}  A_{1j} Y_{1j}} {\sum_{j=1}^{n_1} A_{1j}} $ \\
 &  & $E_0\cup E_1$ & 
    $\frac{1}{M} \sum \limits_{m=1}^M \left\{ \frac{\sum_{j=1}^{n_0}  A_{0j}(1)^{(m)} Y_{0j}(1)^{(m)} + 
 \sum_{j=1}^{n_1}  A_{1j} Y_{1j}
 } {\sum_{j=1}^{n_0} A_{0j}(1)^{(m)} + \sum_{j=1}^{n_1} A_{1j}} \right\}$ \\ \hline
&&&\\ 
 &   & $E_0$ & $\frac{1}{M} \sum\limits_{m=1}^M \left\{ \frac{ \sum_{j=1}^{n_0}  A_{0j} A_{0j}(1)^{(m)} Y_{0j} }
{\sum_{j=1}^{n_0}  A_{0j} A_{0j}(1)^{(m)} } \right\}$ 
\\ 
&  $T=0$ & $E_1$ & $\frac{1}{M} \sum_{m=1}^M \left\{ \frac{ \sum_{j=1}^{n_1} A_{1j} A_{1j}(0)^{(m)} Y_{1j}(0)^{(m)}}
{\sum_{j=1}^{n_1} A_{1j} A_{1j}(0)^{(m)} } \right\}$
\\ 
 &  & $E_0 \cup E_1$ & $\frac{1}{M} \sum\limits_{m=1}^M \left\{ \frac{ \sum_{j=1}^{n_0}  A_{0j} A_{0j}(1)^{(m)} Y_{0j} + \sum_{j=1}^{n_1} A_{1j} A_{1j}(0)^{(m)} Y_{1j}(0)^{(m)} }
{\sum_{j=1}^{n_0}  A_{0j} A_{0j}(1)^{(m)} + \sum_{j=1}^{n_1} A_{1j} A_{1j}(0)^{(m)} } \right\}$ 
\\ \cline{2-4}
\rb{$S_{++}$} &  & $E_0$ & $\frac{1}{M} \sum\limits_{m=1}^M \left\{ \frac{ \sum_{j=1}^{n_0} A_{0j} A_{0j}(1)^{(m)} Y_{0j}(1)^{(m)}}
{\sum_{j=1}^{n_0} A_{0j} A_{0j}(1)^{(m)} } \right\}$ 
\\
& $T=1$ & $E_1$ & $\frac{1}{M} \sum\limits_{m=1}^M \left\{ \frac{ \sum_{j=1}^{n_1}  A_{1j} A_{1j}(0)^{(m)} Y_{1j} }
{\sum_{j=1}^{n_1}  A_{1j} A_{1j}(0)^{(m)} } \right\}$ \\
 &  & $E_0 \cup E_1$ & $\frac{1}{M} \sum\limits_{m=1}^M \left\{ \frac{ \sum_{j=1}^{n_0} A_{0j} A_{0j}(1)^{(m)} Y_{0j}(1)^{(m)} + \sum_{j=1}^{n_1}  A_{1j} A_{1j}(0)^{(m)} Y_{1j} }
{\sum_{j=1}^{n_0} A_{0j} A_{0j}(1)^{(m)}  + \sum_{j=1}^{n_1}  A_{1j} A_{1j}(0)^{(m)} } \right\}$ \\
        \hline\hline 
\multicolumn{4}{p{16cm}}{Abbreviation: PS, principal stratum; $E_i (i=0,1)$ is the subset of patients randomized to treatment $i$. } 
\end{tabular}
\label{tab:estimators}
\end{table}

The SAS program implementing the MI-based AdACE estimtor is provided in the supplemental material.

\section{Simulations}

%{\color{red} Table 2 and Table 3 are the format for desired simulation results, and the numbers need to be replaced with the new simulation results.
%Do we need to use Table 2, or have another table (Table 3)? Notes:

%1. For multiple imputations, we need to specify the class variable and imputation method. See presentation slides 49-53 from Ilya's ICSA presentation.

%2. We can use Rubin's method to estimate the variance for each treatment group and the treatment difference. 
%}
In this section, we consider a two-arm, parallel, randomized trial in diabetes with the simulation settings as described in Qu et al. \cite{qu2020general}. The simulated data are denoted by $(X_{j}, \bd Z_{j}, Y_{j}, \bd I_{j})$ for subject $j$, where $Y_{j}$ is the primary outcome of HbA1c at Week 24 of treatment, $X_{j}$ is baseline HbA1c, $\bd Z_{j} = (Z_{j}^{(1)}, Z_{j}^{(2)}, Z_{j}^{(3)})'$ is a vector of intermediate repeated measurements of HbA1c reading at Weeks 6, 12, and 18, and $\bd I_{j} = (I_{j}^{(1)}, I_{j}^{(2)}, I_{j}^{(3)})'$ denotes the adherence to treatment after Weeks 6, 12, and 18, respectively. The data are simulated for treatments $T=0,1$, respectively.

The baseline value $X_{j}$, intermediate readings $\bd Z_{j}$, and primary outcome $Y_{j}$ are generated by:
\begin{equation} \label{eq:model_x}
X_{j} \sim NID(\mu_x, \sigma_x^2),
\end{equation}
\begin{equation} \label{eq:model_z}
Z_{j}^{(k)} = \alpha_{0k} + \alpha_{1k} X_{j} + \alpha_{2k} T_{j} + \eta_{j}^{(k)}, \quad  1 \le k \le 3, 
\end{equation}
and:
\begin{equation} \label{eq:model_y}
Y_{j} = \beta_0 + \beta_{1} X_{j} + \beta_{2} T_{j} + \sum_{k=1}^3 \beta_{3k} Z_{j}^{(k)} + \epsilon_{j}, 
\end{equation}
where {\em NID} means normally independently distributed, 
$\eta_{j}^{(k)} \sim NID(0, \sigma_\eta^2)$ and $\epsilon_{j} \sim NID(0, \sigma_\epsilon^2)$,
and  $\eta_{j}^{(k)}$'s and $\epsilon_{j}$ are independent.

%\The randomized treatment code $T$ (determining data from which treatment can be observed) is generated independently from a Bernoulli distribution with probability of 0.5.
%\{\color{red} In (2) and (3), you assumes odd subject index is for experimental treatment and even index for the control, which is contradictory to the above statement. Can you use the notation in our previous article?}

We assume patients can drop out of the study after the collection of clinical data at time point $k$. 
The adherence indicator after time point $k (1\le k \le 3)$ is generated from a logistic model:
\begin{eqnarray} \label{eq:model_A}
\logit\{\Pr(I_{j}^{(k)}=1|I_{j}^{(k-1)}=1, X_j, Z_{j}^{(k)})\} = \gamma_0 +  \gamma_1 X_{j} +  \gamma_{3k} Z_{j}^{(k)}, 
\end{eqnarray}
where $\logit(p)=\log(p/(1-p))$, and by convention we set $I_{j}^{(0)}=1$. 
If the adherence indicator at any time point is 0, then the data after this time point are set to be missing.  

To mimic the response and treatment adherence rates in clinical trials for anti-diabetes treatments, we consider two settings in our simulation and the parameters are given: 
$\mu_x = 8.0$, $\sigma_x = 1.0$, $\alpha_{01} = \alpha_{02} = \alpha_{03}= 2.3$, 
$\alpha_{11} = \alpha_{12} = \alpha_{13} = -0.3$, $\alpha_{21}=-0.4$, $\alpha_{22}=-0.9$, $\alpha_{23}=-1.2$,
$\beta_0=0.2$, $\beta_1=-0.02$, $\beta_2=-0.2$,
$\beta_{31}=0.2$, $\beta_{32}=0.4$, $\beta_{33}=0.7$, $\sigma_\eta=0.4$, $\sigma_\epsilon=0.3$,
$\gamma_0 = 3$, and $\gamma_1=-0.1$ (Setting 1) or -0.25 (Setting 2), $\gamma_{31}=-1$, $\gamma_{32}=-2$, $\gamma_{33}=-2.5$, and $j$=1 to 150. %These simulation parameters are the same for $T=0, 1$, respectively. 
These simulation parameters are selected to mimic a real clinical trial in diabetes.

To impute the potential outcome of $(\bd Z_{ij}(1-i), Y_{ij}(1-i), \bd I_{j}(1-i))$ under treatment $(1-i)$ for patient $j$ assigned in treatment group $i$, we need to use the data of $X_{ij}$ and $(X_{1-i,j}, \bd Z_{1-i,j}, Y_{1-i,j}, \bd I_{1-i,j})$. We first create missing records for $(\bd Z_{ij}(1-i), Y_{ij}(1-i), \bd I_{ij}(1-i))$ and then apply an MI procedure to impute these ``missing" values. Note the true missing values at treatment group $(1-i)$ as a result of dropout are also imputed simultaneously. 

The following three steps are then implemented to impute the data: 1) use regression models to impute $\bd Z_{ij}(1-i) | X_{ij}$ based on the relationship between $\bd Z_{1-i,j}$ and $X_{1-i,j}$, 2) impute $Y_{ij}(1-i) | (X_{ij}, \bd Z_{ij}(i-1))$ per the regression model of $Y_{1-i,j} \sim X_{1-i,j} + \bd Z_{1-i,j}$, and 3) impute  $\bd I_{ij}(1-i) | (X_{ij}, \bd Z_{ij}(1-i))$ based on the relationship between $\bd I_{1-i,j}$ and $X_{1-i,j}$ through multiple logistic regressions. These three steps of multiple imputations can be easily implemented through SAS PROC MI procedure. SAS code for the simulation is provided as a supplementary document.

\renewcommand{\baselinestretch}{1.1}
\begin{table}[h!tb] \centering
\caption{Summary of the simulation results for the estimators of treatment effect in two populations of adherers 
(based on 3,000 simulated samples)}
\begin{tabular}{cccccccccc}
\hline\hline
& & & & & \multicolumn{2}{c}{Bootstrap} && \multicolumn{2}{c}{Rubin} \\
\cline{6-7} \cline{9-10}
\rb{Setting} & \rb{Parameter} & \rb{True} \rb{Value}& \rb{Estimate} & \rb{Bias} & SE & CP && SE & CP \\ 
\hline
& $\mu_{0,*+}$ & -0.102	&	-0.102	&	-0.001	&	0.049	&	0.951	&& 0.052  & 0.965 \\
& $\mu_{1,*+}$ & -1.588	&	-1.587	&	0.001	&	0.046	&	0.940	&& 0.047  & 0.950 \\
& $\mu_{d,*+}$ & -1.487	&	-1.485	&	0.002	&	0.057	&	0.944	&& 0.060  & 0.962 \\
\cline{2-10} 
\rb{1}
& $\mu_{0,++}$ & -0.192	&	-0.191	&	0.001	&	0.052	&	0.949   && 0.058  & 0.971	\\
& $\mu_{1,++}$ & -1.638	&	-1.640	&	-0.002	&	0.050	&	0.944   && 0.058  & 0.979	\\
& $\mu_{d,++}$ & -1.446	&	-1.449	&	-0.003	&	0.057	&	0.945   && 0.065  & 0.973	\\
\hline
& $\mu_{0,*+}$ & -0.107	&	-0.108	&	-0.001	&	0.063	&	0.939   && 0.069  & 0.962	\\
& $\mu_{1,*+}$ & -1.606	&	-1.601	&	0.004	&	0.052	&	0.937   && 0.053  & 0.948	\\
& $\mu_{d,*+}$ & -1.499	&	-1.494	&	0.006	&	0.069	&	0.941   && 0.075  & 0.961	\\
\cline{2-10}
\rb{2}
& $\mu_{0,++}$ & -0.272	&	-0.263	&	0.009	&	0.071	&	0.939   && 0.088  & 0.980	\\
& $\mu_{1,++}$ & -1.679	&	-1.678	&	0.000	&	0.064	&	0.941   && 0.088  & 0.990	\\
& $\mu_{d,++}$ & -1.406	&	-1.415	&	-0.009	&	0.069	&	0.944   && 0.095  & 0.990	\\
\hline\hline
\multicolumn{10}{p{16cm}}{Abbreviations: CP, coverage of probability of the 95\% confidence interval; SE, standard error; 
$\mu_{0,*+}$, population mean for the control group for $S_{*+}$; 
$\mu_{1,*+}$, population mean for the treatment group for $S_{*+}$;
$\mu_{d,*+}$, population mean for the treatment difference between treatment and control groups for $S_{*+}$; 
$\mu_{0,++}$, population mean for the control group for $S_{++}$; 
$\mu_{1,++}$, population mean for the treatment group for $S_{++}$; 
$\mu_{d,++}$, population mean for the treatment difference between treatment and control groups for $S_{++}$.
} 
\end{tabular}
\label{table_simulation}
\end{table}
\renewcommand{\baselinestretch}{1.5}

With imputed data, the estimators can easily be calculated through equations provided in Table \ref{tab:estimators}. The true mean responses for each treatment and the treatment difference for $S_{*+}$ and $S_{++}$ can be calculated by numerical integration as described in Qu et al. \cite{qu2020general}. To adjust for baseline covariates, the set of $E_0\cup E_1$ (all patients) is used in the calculation, where $E_i$ is the set of patients randomized to treatment $i$. The point estimates can be obtained by averaging the mean estimates from multiple imputed samples. One method to estimate the variance is to combine the within- and between-imputation variability by Rubin \cite{rubin1976inference} and Barnard and Rubin \cite{barnard1999miscellanea}; it has been reported that these methods may provide conservative coverage probability \citep{robins2000inference, hughes2016comparison, bartlett2020bootstrap}. Our simulations also demonstrate that the variability achieved through the methods of Barnard and Rubin \cite{barnard1999miscellanea} is too conservative and the coverage probability of the confidence interval is larger than its nominal value.  Therefore, we also use a bootstrap method to estimate the variance of the estimator. The bootstrap samples were first created and the imputation procedure was applied to the bootstrap samples. More discussions on various bootstrap methods can be found in Bartlett and Hughes \cite{ bartlett2020bootstrap, schomaker2018bootstrap}. 

Table \ref{table_simulation} shows the simulation results based on 3,000 simulated samples. For each simulated sample, 200 imputations were implemented based on the multiple imputation procedure described earlier. For the bootstrap approach, 50 bootstrap samples are generated for estimating the variance, and then 95\% confidence intervals are calculated based on a normal approximation, which is appropriate in the simulation since the simulated data are generated from a normal distribution. The reason that the percentile bootstrap was not used for constructing the confidence interval is that it requires a large number of bootstrap samples and is very time consuming. The estimates from both scenarios have little bias and the empirical coverage probability for the 95\% confidence intervals is close to the nominal level with the bootstrap approach. Specifically, by comparing the two scenarios in coverage probability, scenario 2 with lower adherence than scenario 1 has slightly lower but acceptable coverage. The 95\% confidence interval based on Rubin's method has much higher coverage probability than the normal level of 0.95. The much wider confidence interval of Rubin's method is likely due to the uncongeniality between the analysis model and imputation model \cite{bartlett2020bootstrap,meng1994multiple,xie2017dissecting}.

We also compared the performance of the proposed method to previously published principal scores method where the potential outcome and adherence are modeled through baseline covariates only \cite{hayden2005estimator,jo2009use}. Similarly, the principal score based method was implemented using MI. With the same simulation settings, Table \ref{table_comparison} shows that the estimates by the principal scores method were more biased as compared to the proposed method utilizing postbaseline intermediate outcomes. For the principal score method, the bias in the estimator for $\mu_{d,++}$ is very small and the estimator for $\mu_{*+}$ is relatively large.

In addition, the type 1 error for the estimate of treatment difference with the proposed method was assessed by updating $\alpha_{2k}=0$ and $\beta_{2}=0$ without changing other parameters in the two simulation settings. Under this scenario (null scenario), the two treatment groups have the same distributions in all variables ($X$, $Z$, $Y$ and $A$). Under the null scenario, the true treatment difference $\mu_{d,++} = 0$ for the principal stratum $S_{++}$. The simulation showed that the rejection rate for the null hypothesis is 0.0680 and 0.0683 for Setting 1 and Setting 2, respectively.  For principal stratum $S_{*+}$, the treatment effect $\mu_{d,*+}$ under the null scenario is not equal to 0, which can be seen by Equation (B.6) in Qu et al. \cite{qu2020general}. This phenomenon will be discussed further in Section \ref{sec:summary}. Therefore, the type 1 error for the estimator for $\mu_{d,*+}$ was not assessed.

%the principal stratum of $S_{*+}$ since the true treatment difference is not 0 even though treatment difference is 0 for the whole population.

\renewcommand{\baselinestretch}{1.1}
\begin{table}[h!tb] \centering
\caption{The comparison for the estimators of treatment effect of adherers between the proposed method and the principal scores method with MI imputation based on baseline only
(based on 3,000 simulated samples)}
\begin{tabular}{cccccccc}
\hline\hline
& & & \multicolumn{2}{c}{Proposed method} && \multicolumn{2}{c}{Principal scores method} \\
\cline{4-5} \cline{7-8}
\rb{Setting} & \rb{Parameter} & \rb{True} \rb{Value}  & Estimate & Bias && Estimate & Bias \\ 
\hline
& $\mu_{0,*+}$ & -0.102	&	-0.102	&	-0.001	&& -0.154  & -0.052 \\
& $\mu_{1,*+}$ & -1.588	&	-1.587	&	0.001	&& -1.596  & -0.007 \\
& $\mu_{d,*+}$ & -1.487	&	-1.485	&	0.002	&& -1.442  &  0.045 \\
\cline{2-8} 
\rb{1}
& $\mu_{0,++}$ & -0.192	&	-0.191	&	0.001   && -0.217   & -0.026	\\
& $\mu_{1,++}$ & -1.638	&	-1.640	&	-0.002   && -1.662  & -0.024	\\
& $\mu_{d,++}$ & -1.446	&	-1.449	&	-0.003   && -1.444  &  0.002	\\
\hline
& $\mu_{0,*+}$ & -0.107	&	-0.108	&	-0.001   && -0.214  & -0.108	\\
& $\mu_{1,*+}$ & -1.606	&	-1.601	&	0.004   && -1.614  & -0.008	\\
& $\mu_{d,*+}$ & -1.499	&	-1.494	&	0.006   && -1.400  & 0.099	\\
\cline{2-8}
\rb{2}
& $\mu_{0,++}$ & -0.272	&	-0.263	&	0.009   && -0.303  & -0.031	\\
& $\mu_{1,++}$ & -1.679	&	-1.678	&	0.000   && -1.707  & -0.028	\\
& $\mu_{d,++}$ & -1.406	&	-1.415	&	-0.009   && -1.404  & 0.003	\\
\hline\hline
\multicolumn{8}{p{14cm}}{Abbreviations: 
$\mu_{0,*+}$, population mean for the control group for $S_{*+}$; 
$\mu_{1,*+}$, population mean for the treatment group for $S_{*+}$;
$\mu_{d,*+}$, population mean for the treatment difference between treatment and control groups for $S_{*+}$; 
$\mu_{0,++}$, population mean for the control group for $S_{++}$; 
$\mu_{1,++}$, population mean for the treatment group for $S_{++}$; 
$\mu_{d,++}$, population mean for the treatment difference between treatment and control groups for $S_{++}$.
} 
\end{tabular}
\label{table_comparison}
\end{table}
\renewcommand{\baselinestretch}{1.5}

\section{Application}
\label{sec:application}
The application of the proposed method was based on the IMAGINE-3 Study, which has been used by Bergenstal et al. \cite{bergenstal2016randomized} and allows for direct comparison with previous results. IMAGINE-3 was a 52-week treatment trial for patients with type 1 diabetes mellitus to demonstrate basal insulin lispro (BIL) was superior to insulin glargine (GL). In this trial, 1,114 adults with type 1 diabetes were randomized to BIL and GL in a 3:2 ratio.  The study was conducted in accordance with the International Conference on Harmonisation Guidelines for Good Clinical Practice and the Declaration of Helsinki. This study was registered at clinicaltrials.gov as NCT01454284 and details of the study report have been published.

Of the 1,114 randomized patients, 1,112 patients (663 in BIL, 449 in GL) took at least one dose of study drugs. A total of 235 patients permanently discontinued the study treatment early due to reasons of lack of efficacy (LoE), adverse events (AE), or adminstration reasons, leaving 877 (78.9\%) patients adhering to the treatment.

To apply to the proposed methods, we consider the following baseline covariates $\bd X$ that could potentially impact treatment adherence: age, gender, HbA1c, low density lipoprotein clolesterol (LDL-C), triglyceride (TG), fasting serum glucose (FSG), and alanine aminotransferase (ALT). The study also collected HbA1c, LDL-C, TG, FSG, and ALT at Week 12 and Week 26, and injection site reaction adverse events (a binary variable) that occurred between randomization and Week 12 and between Week 12 and Week 26. Those post baseline variables were considered in intermediate covariates $\bd Z_1$ for Week 12 and $\bd Z_2$ for Week 26, respectively. The primary outcome $Y$ was the HbA1c reading at Week 52.

For each stratum of $S_{*+}$ or $S_{++}$, 1,000 imputations were generated and the complete data after imputations were used to estimate the mean response for each treatment group and the treatment difference. Due to a large amount of ``missing" values (potential outcomes for the alternative treatments were not observed), we need a large number of imputations to achieve good accuracy for the estimates. Based on our investigation, the 1,000 imputed samples made the variance due to imputation random error $<0.1\%$ of the variance of the final estimate (average of the estimates from 1,000 imputations).  The variance of the final estimate was estimated using 50 bootstrap samples and the corresponding 95\% confidence interval was calculated using the normal approximation with the bootstrap variance. Due to the relatively large sample size, we expect the distributions of the estimators to be approximately Gaussian, which was confirmed by normal Q-Q plots of the 50 bootstrap estimates for all parameters for $S_{*+}$ and $S_{++}$ (data not shown here).

\begin{table}[h!tb]
\centering
\caption{\small Summary of results of the real data analysis for the estimators of treatment effect in HbA1c for the two populations of adherers using proposed methods}
\label{table_example}
\small
    \begin{tabular}{cccccc}
    \toprule
  \multirow{2}[1]{*}{Treatment} & \multicolumn{2}{c}{$S_{*+}$} & & \multicolumn{2}{c}{$S_{++}$} \\
   \cline{2-3} \cline{5-6}
  & Estimate $\pm$ SE & 95\% CI & & Estimate  $\pm$ SE & 95\% CI \\
   \cline{1-6}
 GL   & 7.59 $\pm$ 0.05 & (7.49 , 7.70) & & 7.54 $\pm$ 0.05 & (7.44 , 7.64) \\
BIL    & 7.33 $\pm$ 0.04 & (7.25 , 7.41) && 7.30 $\pm$ 0.04 & (7.22 , 7.37) \\
 BIL vs. GL & -0.26 $\pm$ 0.05 & (-0.35 , -0.16) && -0.25 $\pm$ 0.05 & (-0.34 , -0.15) \\
      \bottomrule
\multicolumn{6}{p{15cm}}{Abbreviations: BIL, basal insulin peglispro; CI, confidence interval; GL, basal insulin glargine; SE, standard error} 
\end{tabular}
\end{table}

Table \ref{table_example} shows the estimates for HbA1c at 52 weeks for each treatment group and the treatment difference for the population $S_{*+}$ and $S_{++}$ using the method based on $E_0 \cup E_1$. The estimates were similar to those reported in Qu et al. \cite{qu2021implementation}. It showed BIL was superior to GL in controlling HbA1c for the two principal strata: $S_{*+}$ and $S_{++}$. 

\section{Summary and Discussion} \label{sec:summary}
In addition to the commonly used estimands for the treatment difference for the whole study population, the treatment difference for adherers (a principal stratum) is also important and plays a primary role in assessing the effect of a treatment as described in the so-called tripartite approach \citep{ ruberg2017considerations, akacha2017estimands}. When making decisions whether to start a new pharmacologic treatment or not, patients and physicians want to know what the effects of that treatment are when the patient takes the medication as prescribed. Careful thought and more sophisticated analyses are required (i.e., not the naïve completers analysis) so that data from randomized clinical trials can be used to assess this important principal stratum and provide an estimate of the causal effect of the treatment. 

Furthermore, some aspects of clinical practice remain trial and error; a patient is prescribed a medication and follow-up visits are scheduled to assess the status of the patient’s disease or any resulting side effects. Those observations are used to guide the patient’s subsequent treatment with dosage changes or switching to other treatments. In our interactions with physicians, many start by prescribing a treatment that is highly effective when taken at the recommended dose and frequency and only alter or discontinue that treatment if side effects are not tolerable or if there is insufficient efficacy. This is considered preferable to starting with a treatment of lesser efficacy but perhaps greater adherence due to fewer or more acceptable side effects. Such treatments can always be a “fallback” option. 

These considerations suggest that the treatment effect estimate in the principal stratum of patients who can adhere to treatment is very important, if not more important than the estimate for the whole study population. Furthermore, the Tripartite Approach of Akacha et al. \cite{akacha2017estimands} is ideally equipped to provide not only an estimate of the treatment effect in this clinically meaningful principal stratum, but also the likelihood of non-adherence due to adverse events and lack of efficacy. With this information, the patient and their physician can have a meaningful and nuanced discussion about the expected benefits and risks of actually taking the medication.

Finally, side effects are most often analyzed and described in the context of what happens when a medication is taken as prescribed, and we believe this context is most relevant for efficacy as well, especially when assimilating information into benefit-risk assessments.

Qu et al. \cite{qu2020general} and Zhang et al. \cite{zhang2021statistical} provide the general framework for estimating the AdACE, but the implementation of such estimators is rather complex. In this article, we proposed an MI-based method to construct the estimators, which is much more straightforward than the original method proposed to construct the AdACE estimators. We evaluated the performance through simulations and it showed that the new method provides consistent estimators and has the correct coverage probability for the bootstrap confidence interval at the nominal alpha level. We also applied these MI-based estimators to the same data set as in Qu et al. \cite{qu2021implementation} and yielded similar results as the original estimates. 

We have focused on the discussion of the proposed method for the estimation of the mean treatment difference for a continuous variable. This method can be easily adapted to different types of variables (e.g., binary or time-to-event variables) as long as the data can be imputed consistently. Imputation of data for binary and time-to-event variables, which may be more challenging, is out of the scope of this article.

The proposed method, which utilizes the intermediate outcomes, is certainly more complex than the traditional method of identifying principal strata or predicting the outcome measurement only using baseline covariates. When the adherence only depends on baseline covariates, the proposed method will not provide additional benefit compared to the method of modeling the principal score only through the baseline covariates, but it will generally not deteriorate the estimation unless too many intermediate outcomes are included. In practice, baseline covariates and intermediate outcome variables should be carefully selected. In the example in Section \ref{sec:application}, the selected variables were a direct indication of efficacy and safety outcomes that could potentially lead to treatment discontinuation.  

It requires a set of relatively strong assumptions (A1-A7) to make the proposed estimator based on multiple imputation consistent for the treatment effect for adherers. Therefore, sensitivity analyses may be performed. The sensitivity analyses can be easily performed within the imputation (e.g., adding a sensitivity parameter after the imputation as in the tipping-point analysis \cite{yan2009missing}). Future research on sensitivity analyses may be required. 

It requires careful consideration for using an estimand for a principal stratum. In general settings (e.g., in the setting in this article), the true treatment effect for $S_{*+}$ under the null treatment effect case may not zero. For the simulation model in this article, this can be seen by the theoretic treatment difference for $S_{*+}$  derived in Equation (B.6) in Qu et al. \cite{qu2020general}. 
Only under special cases, the true treatment effect for $S_{*+}$ is equal to zero under the null hypothesis of no treatment difference between the potential outcomes for the 2 treatments. Further research is required to find the necessary conditions for which $\mu_{d,*+} = 0$ under the null hypothesis. This phenomenon provides another reason to use the treatment effect for $S_{++}$, in which the treatment effect is always zero under the null hypothesis. In addition, the traditional method of forming the null hypothesis may not work for principal stratification based estimands and the principal stratification variables should be take into consideration. 

In summary, the MI-based estimation proposed in this article will allow for broader adoption and easier estimation of the AdACE, providing estimation for an alternative clinically meaningful estimand for adherers.

\section*{Acknowledgements}
We would like to thank Dr. Ilya Lipkovich for his scientific review and useful comments. We would also like to thank the Associate Editor and two anonymous referees for their valuable comments, which lead to significant improvement of this article.  

\section*{Conflict of interest}
There is no funding received for this research other than the time spent as part of employment. For Junxiang Luo, the work was done when he was an employee of Sanofi. 

\section*{Data sharing}
Authors elect to not share data. 

\bibliographystyle{vancouver}
\bibliography{references}

\end{document}